\def\cm{cm$^{-1}$}
\documentstyle[prl,aps,graphicx,multicol]{revtex}
\begin{document}
\draft
\title{Competition between Charge Ordering and Superconductivity\\ 
in Layered Organic Conductors
$\alpha$-(BEDT-TTF)$_2M$Hg(SCN)$_4$ ($M$ = K, NH$_4$) }
\author{M. Dressel$^{1}$\cite{email}, N. Drichko$^{1,2}$, J. 
Schlueter$^3$, and J. Merino$^4$}
\address{$^1$ 1.~Physikalisches Institut, Universit\"at Stuttgart,
Pfaffenwaldring 57, D-70550 Stuttgart, Germany\\
$^2$ Ioffe Physico-Technical Institute,  St.\ Petersburg, Russia\\
$^3$ Material Science Divisions, Argonne National Laboratory, 
Argonne, IL 60439-4831\\
$^4$ Max-Planck-Institut f\"{u}r Festk\"{o}rperforschung, D-70506 Stuttgart, 
Germany\\
}
\date{Received  \today}
\maketitle
\begin{abstract}
While the optical properties of the superconducting salt
$\alpha$-(BEDT-TTF)$_2$\-NH$_4$Hg(SCN)$_4$ 
remain metallic down to 2~K, in the non-superconducting 
K-analog a pseudogap develops 
at frequencies of about 200~cm$^{-1}$ for temperatures
$T < 200$~K. Based on exact
diagonalisation calculations on an extended Hubbard model at
quarter-filling we argue that fluctuations associated
with short range charge ordering are responsible 
for the observed low-frequency feature.
The different ground states, including superconductivity, 
are a consequence of the proximity of these compounds
to a quantum phase charge-ordering transition driven by the 
intermolecular Coulomb repulsion. 
\end{abstract}
\pacs{PACS numbers: 74.70.Kn, 78.30.Jw, 71.27.+a}

\begin{multicols}{2}
\columnseprule 0pt 
\narrowtext
Electronic and magnetic systems in low dimensions display 
a number of exciting phenomena which have 
been the subject of many experimental and theoretical 
investigations \cite{Auerbach}.
Ideal realizations of such systems are the organic conductors 
in which the effect of strong electronic correlations in the
ground state properties is under intense investigation\cite{Ishiguro98}. 
For example, in the one-dimensional Bechgaard salts TMTSF and TMTTF,  
superconductivity, charge ordering, and spin-density-wave 
instabilities are observed below a metallic
state (which is described by a Fermi-liquid or a Luttinger liquid, depending on the strength of the interchain coupling)\cite{Ishiguro98,Vescoli98}.
Significant progress has been achieved in understanding their properties 
in comparison to the quasi two-dimensional materials,
for which the theoretical models and experimental studies
are much less advanced. While some members of the BEDT-TTF family 
(where BEDT-TTF or ET stands for
bis\-ethylene\-di\-thio\--tetra\-thia\-fulvalene), which serves as a model system 
for layered conductors, become
superconducting, others remain metallic down to the lowest
temperatures accessible, and others undergo metal-to-insulator
transitions \cite{McKenzie98}. This is in contrast with band structure
calculations which predict a metallic state for most of these salts.

Besides the well-studied $\kappa$-phase compounds, 
the $\alpha$-phase family of
(BE\-DT\--TTF)$_{2}$\-$M$\-Hg\-(SCN)$_{4}$ has attracted
considerable attention after superconductivity at
$T_c=1$~K was found in the NH$_4$-salt \cite{Osada90}. In contrast, the $M$=
K, Rb, and Tl-salts, remain metallic down to a few mK entering
a  density-wave  ground state at $T_p \approx 8$~K to 12 K
which is highly controversial \cite{Brooks95,McKenzie97,Christ00}.
It was suggested that a low-temperature
reconstruction of the Fermi surface induced by a Peierls
distortion \cite{Harrison99} occurs in $M$ = K,  Rb, and 
Tl-salts but not in the NH$_4$-compound. However, the Fermi 
surfaces calculated for all of these compounds 
are almost identical, consisting of two-dimensional closed
pockets\cite{remark1} and quasi-one-dimensional open sheets. This
topology was confirmed by a number of quantum-oscillation
studies\cite{Wosnitza96}. Hence, the different ground states encountered
are difficult to understand from band structure calculations. 
In the present Letter, we introduce a scenario based on 
comprehensive optical investigations combined with exact diagonalisation 
calculations, in which electronic correlations play an essential role.
We show that the proximity of these materials to a quantum phase
transition driven by the intermolecular Coulomb repulsion
can explain the different ground states found in the various salts.

Optics has proven to be a powerful method to explore the
ground state of low-dimensional metals \cite{DresselGruner02}. 
In K$_{0.3}$\-MoO$_3$
or TTF-TCNQ, for example, clear signatures of a charge
density wave have been observed in the
optical spectra, like the opening of a single particle gap  in the infrared
and collective excitation in radio-frequency and microwave range \cite{Gruner94b}.
Similar fingerprints (but at lower energies)
of a spin-density-wave ground state have been
identified in (TMTSF)$_2$PF$_6$ \cite{DresselGruner02}.
Recently the development of a pseudogap in the 
two-dimensional metal (BEDT-TTF)$_4$\-[Ni(dto)$_2$]
was related to the strong Coulomb interaction, and we predicted that it
should also play an important role in the $\alpha$-type compounds \cite{Haas00}.

\begin{figure}
\includegraphics[width=130mm]{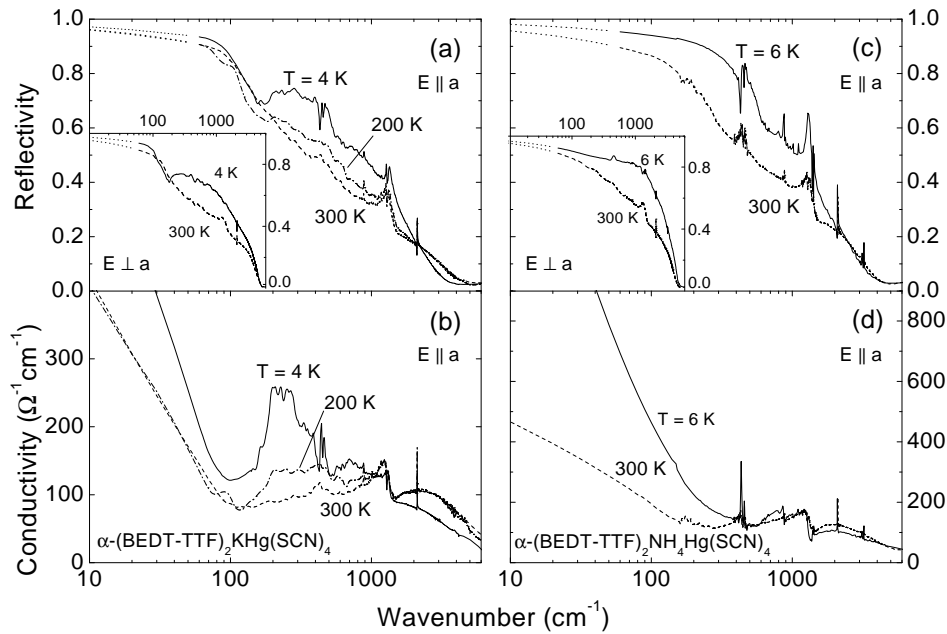}
\caption{\label{fig1} (a) Frequency dependent reflectivity and (b)
optical conductivity of $\alpha$-(BEDT-TTF)$_2$KHg(SCN)$_4$ for
different temperatures as indicated. The panels (c)  and (d) show
the reflectivity and conductivity, respectively, of
$\alpha$-(BEDT\--TTF)$_2$NH$_4$Hg(SCN)$_4$. The measurements were
performed for $E\parallel a$; the $E\perp a$ reflection data are
displayed in the corresponding insets. For both polarization the
reflectivity of (BEDT-TTF)$_2$KHg(SCN)$_4$ has a dip at around 
200~cm$^{-1}$.}
\end{figure}

Single crystals of $\alpha$-(BEDT-TTF)$_2$KHg(SCN)$_4$  and
$\alpha$-(BEDT-TTF)$_2$NH$_4$Hg(SCN)$_4$ (denoted by
K-salt and NH$_4$-salt) of about $1.5\times 1 \times 0.3~{\rm mm}^3$ in size
were studied by polarized optical reflection methods in the
frequency range from 50 to 10\,000~\cm\ at $4~{\rm K}<T<300$~K.
Figs.\ \ref{fig1}a and c show the reflectivity spectra $R(\omega)$ of both
compounds obtained for the electric field $E\parallel a$ 
($a$ corresponds to the stacking direction) at
different temperatures. In Ref.~\onlinecite{Dressel92,Drichko02} 
we have analyzed the vibrational features of both salts in detail. 
Common to many organic conductors, the room-temperature
reflectivity decreases gradually for increasing frequency with no
pronounced plasma edge \cite{Dressel00}. Going  down to low
temperatures, an overall rise in reflectivity is observed in
the NH$_4$-salt, but no drastic changes of the general behavior;
this is expected for a metal. While at frequencies above approximately
500~\cm\ the temperature dependences of both compounds are similar, 
in the K-salt a new feature is observed at lower energies: a
significant dip in the reflection centered around 200~\cm\
gradually develops as $T$ drops below 200~K
(Fig.~\ref{fig1}a). Perpendicular to the $a$-axis the reflectivity
of both materials exhibits a plasma edge around 5000~\cm\ which
becomes more pronounced with decreasing $T$ (insets of
Fig.\ref{fig1}). Again, no significant changes are
observed with lowering $T$ for the NH$_4$-salt; the
K-compound, however,  exhibits a dip in the reflectivity around
200~\cm\ for $T\leq 200$~K which is identical to the other
polarization.

For the Kramers-Kronig analysis
\cite{DresselGruner02} we used a Hagen-Rubens extrapolation for
$\omega\rightarrow 0$; above 7000~\cm\ we utilized the standard
optical behavior known from other BEDT-TTF compounds and finally
extrapolated by $R\propto\omega^{-4}$. The corresponding
conductivity $\sigma(\omega)$ is plotted in the lower frames of
Fig.~\ref{fig1}. The two compounds exhibit a similar room
temperature behavior, only the absolute values of the NH$_4$-salt
are slightly higher. For both polarizations their conductivity 
shows a Drude-like peak and a broad maximum near 2000~\cm.
The shift of the spectral weight to lower energies as $T$ decreases 
agrees with the dc resistivity\cite{remark2}.
In the mid-infrared range the absolute values of
both directions differ by a factor of 2. At $T = 300$~K the
overall shape, however, is not so much different for the two
polarizations in spite of the anisotropic Fermi surface; only when
the temperature is lowered  these differences  become stronger.
The dip in $R(\omega)$ observed in the  K-salt in both
polarizations shows up as a maximum in the conductivity spectra
slightly above 200~\cm\ due to  excitations across the  pseudogap.
The position of the peak is the same
for both orientations and does not change with temperature; the
spectral weight increases linearly by about a factor of 5 when
going from $T=200$~K to 4~K \cite{remark2}; no signs of a phase transition
around $T_p\approx 8$~K are observed in our spectra. In agreement
with previous studies \cite{Dressel92} we do not find indications
of a pseudogap in the NH$_4$-compound.

\begin{figure}
\includegraphics[width=70mm]{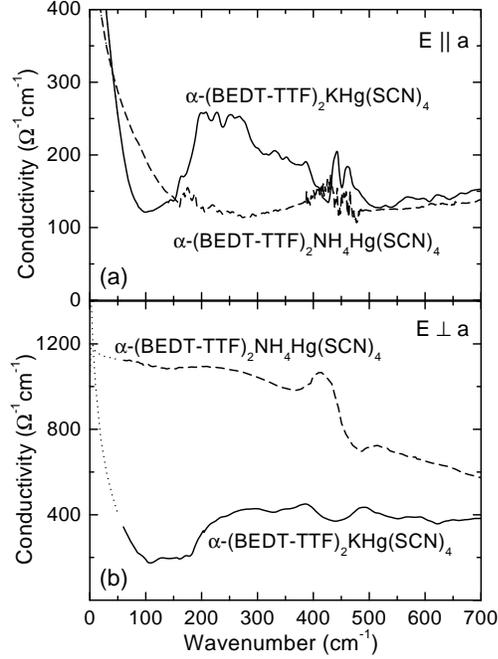}
\caption{\label{fig2}Optical conductivity of 
$\alpha$-(BEDT\--TTF)$_2$\-KHg\-(SCN)$_4$ (solid lines) and 
$\alpha$-(BEDT\--TTF)$_2$\-NH$_4$Hg(SCN)$_4$ (dashed lines) 
obtained at
$T=4$~K for the electric field polarized (a) parallel and (b) perpendicular to 
the $a$-axis.
In both directions of (BEDT-TTF)$_2$KHg(SCN)$_4$ a pseudogap feature is 
clearly seen at 200~cm$^{-1}$
which is not present in the superconductor (BEDT-TTF)$_2$NH$_4$Hg(SCN)$_4$.}
\end{figure}

In Fig.~\ref{fig2}
the low-temperature conductivity of the two materials is compared
for both polarizations. The strong feature is
clearly seen in the spectra of the K-salt at around 200~\cm\ where
$\sigma(\omega)$ drops by 50\%. A narrow Drude-like contribution
with a scattering rate of less than 30~\cm\ remains which contains
only a few percent of the spectral weight; nevertheless it is
responsible for the dc conductivity and the quantum oscillations
\cite{Wosnitza96}. The  overall scenarios are similar in both
directions.

In the last decade a large number of investigations have been performed
on the low-temperature phase of $\alpha$-(BEDT-TTF)$_2$KHg(SCN)$_4$. 
The transition at $T_p\approx 8$~K 
is associated with a change in the Fermi surface
\cite{Sasaki94,Caulfield95} which could open a charge gap along
the $a$-direction; of course large parts of the Fermi surface remain
intact leading to the metallic dc conductivity and quantum
oscillations. A small modulation of the magnetic moments suggest a
spin-density-wave ground state \cite{Sasaki91,Pratt95}; but also a
charge-density-wave instability was recently suggested
\cite{McKenzie97,Christ00,Biskup98}. The magnetic field dependence
of the transition temperature is controversial
\cite{Brooks95,Sasaki96}. Above a so-called kink-field $B_k\approx 24$~T 
the metallic state is restored \cite{Brooks95} but
maybe a new phase emerges \cite{Biskup98,Kartsovnik97}. 
The interpretation of the measured optical spectra of 
$\alpha$-type BEDT-TTF salts in terms of a charge-density-wave 
instability, faces several difficulties:
(i)~the strong $T$ dependence
of the 200~cm$^{-1}$ feature between 4~K and 200~K,
is incompatible with semiconductor-like excitations.
(ii)~The negligibly small anisotropy in the optical response seems
inconsistent with the large anisotropy of the crystal structure,
(iii)~The dip in the reflectivity of the K-salt can be
observed up to $T\approx 200$~K, which is difficult to
associate with the low-temperature density-wave ground state.
In order to overcome the above difficulties, we find it
necessary to consider both the
{\it intramolecular} Coulomb repulsion $U$, as well as the {\it intermolecular} Coulomb
interaction $V$. Indeed, we show that the $V$-term is responsible
for the subtle competition between the different ground states 
appearing in $\alpha$-(BEDT-TTF)$_2\-M$Hg\-(SCN)$_4$.

\begin{figure}
\includegraphics[width=75mm]{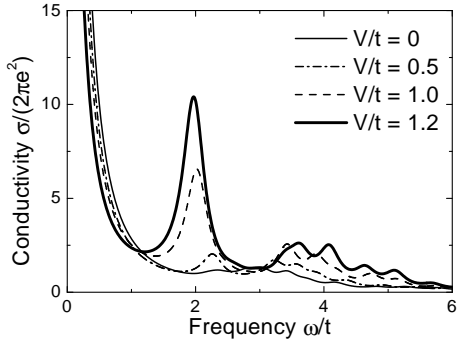}
\caption{\label{fig3}Evolution of optical conductivity as the system is brought
closer to the charge ordering transition. 
The optical conductivity obtained using exact diagonalisation
of an extended Hubbard model on a $4 \times 4$ square lattice with fixed $U=20t$ 
is plotted 
for
increasing intersite Coulomb repulsion $V$. A strong feature develops at low 
frequencies associated with fluctuations due to short range charge ordering.
Note that for these values of $V$ the system is well in the metallic side 
of the transition: $V < V_c^{\rm MI}\approx 2.2 t$. 
}
\end{figure}

It is known that Coulomb interaction leads to
unconventional behavior of the electronic properties of layered
organic materials\cite{McKenzie98,Dressel00}. 
For instance, the optical conductivity of
$\kappa$-type compounds display a Drude peak at low temperatures
which is suppressed for $T>50$~K, in contrast to
conventional metals \cite{Kornelsen89}. This is in
agreement with dynamical mean-field-theory
calculations\cite{Georges96} on a frustrated lattice at
half-filling with strong on-site Coulomb repulsion $U \approx W$
($W$ being the bandwidth). 
Since in $\kappa$-salts the BEDT-TTF molecules are dimerised, 
the conduction band is half-filled. 
The quasi-particles are destroyed with increasing $T$ because the system 
is close to a Mott metal-insulator transition.
This is not the case in the $\alpha$-salts
as they are quarter-filled with holes. For instance, Hartree-Fock
calculations in a Hubbard model with $U=0.7$~eV suggest a
paramagnetic metallic ground state for the K-compound
\cite{Kino95}. Including the nearest-neighbor interaction
$V$, however, can lead to charge-ordering phenomena \cite{Seo00}. 
Recent exact-diagonalisation calculations on an extended Hubbard 
model at quarter-filling \cite{Calandra02} show that a transition from a
metal to a checkerboard charge ordered insulator
occurs at $V=V_c^{\rm MI}=2.2 t$, where $t$ is the nearest-neighbors 
hopping. Also there is a strong redistribution of the
spectral weight in the optical conductivity.
Calculations of the optical conductivity on a $4\times 4$ 
square lattice are shown in Fig.~\ref{fig3} 
for $U=20 t$ and different values of $V < V_c^{\rm MI}$.
We find a broad resonance and a sharp peak located at frequencies of 
about 2t. When the system is insulating, the calculated spectra 
consists of a single broad resonance centered at $\omega  \approx 3V$; 
this is the energy cost for moving
an electron inside the checkerboard. However, for small values of $V$   
the checkerboard is not fully formed, the energy cost
being smaller than $3V$. 
We associate this broad resonance resulting from the
calculations shown in Fig.~\ref{fig3} with the 2000~cm$^{-1}$ band
experimentally observed. On the other hand, we attribute the low frequency feature 
to the enhancement of charge fluctuations associated with short range 
charge ordering. We note that as $V/t$ increases this resonance shifts down slightly
and the intensities of both the broad band and the sharp peak are
enhanced. For $V \leq 0.5 t$ the spectrum is dominated by the
Drude peak, in agreement with the optical conductivity observed for the NH$_4$-salt.
Note that the strong enhancement of the sharp feature  
occurs already at $V \approx t$. Our calculations are in qualitative good 
agreement with the evolution of the measured optical spectra when going from
the NH$_4$-salt to the K-analog if we assume that $V/t$ increases
from the former to the latter. We note that the 
Fermi surface considered in the calculations does not have any special property 
(for the quarter-filled square it consists of a featureless closed sheet only),
making our results robust against changes in its shape.
 
In order to explain why superconductivity occurs 
in the NH$_4$-salt but not in the corresponding K-compound, 
we consider results from many-body 
calculations on an extended Hubbard model at quarter filling. 
Slave-boson theory predicts the appearance of
superconductivity mediated by charge fluctuations which are 
present on the metallic side of the phase diagram close to the charge-ordering
transition\cite{Merino01}. This suggests that superconductivity 
in NH$_4$ is a consequence of its proximity to a charge-ordering instability
driven by the ratio $V/t$. Hence,
$\alpha$-(BEDT-TTF)$_2$KHg(SCN)$_4$, which is at the
charge-ordered side of the transition, {\it i.e.} $(V/t)_{\rm K}>(V/t)_{\rm NH_4}$, 
may be converted into a superconductor by decreasing $V/t$.
Some attempts in this
direction have already been carried out by applying  external
pressure\cite{Brooks95}. 
Uniaxial strain provides a better way to tune the
materials through the transition because decreasing $a/b$ 
(where $a$ is the intra-stack and $b$ the inter-stack distances)
favors a metallic state.
As a matter of fact, $T_c$ increases from 1.5~K up to about 6~K
in the NH$_4$-compound; and most important, superconductivity is 
also reached for the K-salt when sufficient pressure 
is applied along the $a$-direction \cite{Maesato01}.
The above scenario is consistent with 
superconductivity at quantum phase transitions 
observed in numerous systems\cite{McKenzie98}. 
Hence the NH$_4$-salt can be driven closer to the 
charge-ordering transition by increasing $V/t$;
superconductivity should disappear, and eventually we expect
a redistribution of the spectral weight due to the gradual
enhancement of dynamically induced charge fluctuations.
In contrast to the K-compound, for the superconducting NH$_4$-salt
the Coulomb repulsion is not strong 
enough to induce charge ordering. 
Releasing the pressure sufficiently should suppress superconductivity
and eventually lead to a charge-ordered insulator.

In conclusion, we have investigated the optical properties of the
metallic ($M$ = K) and superconducting ($M$ = NH$_4$) compounds of
$\alpha$-(BEDT-TTF)$_2M$Hg(SCN)$_4$. Due to the
closer proximity to a charge-ordering quantum phase transition, the
K-salt shows a strong feature in the electronic spectrum at
about 200~\cm\ while  the optical conductivity of the
NH$_4$-compound remains metallic at all temperatures until
the system eventually becomes superconducting. From exact 
diagonalisation calculations we identify the low-frequency
feature with the gradual enhancement of fluctuations
associated with short range ordering close to a  
charge-ordering transition. This scenario may be tested
by applying uniaxial strain to tune the materials 
through the transition and induce superconductivity in the K-salt.
Further experiments using X-ray and Raman scattering are necessary to 
provide additional evidence of charge ordering phenomena in the $\alpha$-salts.                                                

We thank F. Assaad, M. Calandra, M. Kartsovnik, R. H. McKenzie,
H. Seo and R. Zeyher for helpful discussions. The work at Argonne
National Laboratory was performed under the auspices of the Office
of Basic Energy Sciences, Division of Material Sciences of the
U.S. Department of Energy, Contract W-31-109-ENG-38. N.D
acknowledges support from the DAAD, Germany. J. M. was supported
by a Marie Curie Fellowship of the European Community program
``Improving Human Potential'' under contract No. HPMF-CT-2000-00870.

\end{multicols}

\newpage

\begin{thebibliography}{99}
\bibitem{remark2}In all cases the overall spectral weight is conserved.
\end{thebibliography}
\end{document}